\documentclass[a4paper]{llncs}
\usepackage[utf8]{inputenc}
\usepackage[T1]{fontenc}
\usepackage[english]{babel}
\usepackage{amsmath}
\usepackage{amssymb,amsfonts,textcomp}
\usepackage{array}
\usepackage{hhline}
\usepackage{graphicx}
\usepackage{cite}

\usepackage{url}
\urldef{\mailsa}\path|{Trenton.Schulz,Kristin.Skeide.Fuglerud}@nr.no|
\newcommand{\keywords}[1]{\par\addvspace\baselineskip
\noindent\keywordname\enspace\ignorespaces#1}

\begin{document}
\mainmatter
\title{Creating Personas with Disabilities}
\author{Trenton Schulz \and Kristin Skeide Fuglerud}
\institute{Norsk Regnesentral -- Norwegian Computing Center,\\
Gaustadalléen 23a/b, Kristen Nygaards hus, NO-0373 Oslo, Norway\\
\mailsa\\
\url{http://www.nr.no}}
\toctitle{Creating Personas with Disabilities}
\tocauthor{Trenton Schulz}
\maketitle

\begin{abstract}
    Personas can help raise awareness among stakeholders about users' needs. While
    personas are made-up people, they are based on facts gathered from user research.
    Personas can also be used to raise awareness of universal design and accessibility
    needs of people with disabilities. We review the current state of the art of the
    personas and review some research and industry projects that use them.  We outline techniques that can be used to create
    personas with disabilities. This includes advice on how to get more information about
    assistive technology and how to better include people with disabilities in the persona
    creation process. We also describe our use of personas with disabilities in several
    projects and discuss how it has helped to find accessibility issues.
\end{abstract}

\section{Introduction}
Personas are a great way of raising awareness of user needs throughout a project. A
persona is a rich description of a potential user of your system and consists of several
stereotypical traits, such as preferences, needs, attitudes, habits, and desires, composed
into a realistic, but fake, person with a name and picture. Instead of arguing for the
needs of a generic user that can morph from being a complete novice in one situation to an
experienced expert in another, stakeholders ground themselves in the facts of the personas in
front of them. The idea is that the resulting system will be more successful when it is
designed with specific users in mind rather than a vague idea of an average user.

Since personas help focus on the concrete facts about your potential
users, it would be beneficial if the needs of people with disabilities are included among
the personas. While this has been encouraged and used in different places, there is little
information on how one should create personas with disabilities. This results in a risk of
creating personas that do not capture the needs of people with disabilities or that are
based on incorrect information. We introduce a methodology for creating personas with
disabilities that minimizes this risk by providing ways of collecting information and
opinions from the people with disabilities and taking into account the different Assistive
Technology (AT) they use. This methodology can be easily included in a standard persona
creation methodology contributing to systems that are designed for all.

\section[State of the Art]{State of the Art}
\label{SOTA}
Input from users is important to ensure that new systems can be usable and cover their
needs. Participation from the users is one of the principles behind Scandinavian Design
\cite{Ehn1993}. Yet, having constant access to users is resource-intensive and may be too
costly for a project. Using personas helps ensure that users' perspectives are included.
Lindgren, Chen, Amdahl, and Chaikiat \cite[p.~461]{Lindgren2007} describe personas as
``\ldots a hypothetical archetype of real users described in great detail and defined by
their goals and needs, rather than just demographics.'' Personas are usually presented in
a persona description, this usually includes a picture of the persona; background
information including family, tasks, motivations, attitudes to technology and technical
knowledge.

Personas were popularized by Cooper \cite{Cooper1999} in his book \emph{The Inmates are Running the Asylum}.
Cooper presents personas as a way to include viewpoints from different user groups without
falling into the trap of using a generic user. He stresses that designing something for
specific people will be more successful than trying to create something that works for
generic users. Cooper then uses the persona technique to aid in designing an entertainment
system for an airline. While one of the personas is an elderly man with some vision
problems, there is little information about how Cooper created these personas.

Personas are normally used to keep the focus on the users on the project, but it is also
possible to find out about your users based on the personas you have created. Chapman,
Love, Milham, ElRif, and Alford \cite{Chapman2008} have demonstrated a novel use of
personas. They took personas' properties and found how prevalent each property was among their user groups.

To ensure that systems we design are accessible, we should also create personas that have
disabilities. Zimmermann and Vanderheiden \cite{Zimmermann2007} point out that using
personas with impairments help make the accessibility requirements real. Using personas
with disabilities gives us an opportunity to include their needs without the resource
intensive task of recruiting disabled people for all stages of the project. Although he
does not provide detailed instructions on how to create personas, Henry \cite{Henry2007}
encourages the use of personas with disabilities in the design of ICT, but reminds us that
everything that is true for one person with one disability is not necessarily applicable
to all people with that disability. This advice, however, is useful for all kinds of
personas.

The use of personas with disabilities has gained traction in the industry and in several
recent EU research projects like ACCESSIBLE \cite{Isacker2008a} and ÆGIS \cite{Isacker2008}. The Ubuntu operating system
has also adapted personas in guiding its accessibility development \cite{Bell2011}. These projects
have provided their personas online. Others have encouraged people to use these specific
personas for other projects \cite{Korn2010}. This may seem like a shortcut for creation, but recycling
personas is not recommended \cite{Pruitt2006}. This is because an equally important aspect
is to engage the stakeholders and the development team, to let them get to know the
personas and to empathize with them. This part is lost when recycling personas from
another project. Knowledge of creating personas can be recycled, however.

Pruitt and Grudin \cite{Pruitt2003} list several problems with personas in projects. The personas are not believable, because they were not based on real data; that they are not presented well, and then not used; no understanding of how to use the personas; and finally that only part of the group is interested in using personas and there are no resources to make personas come alive. There are several methods for combating these problems and we will present some in Section \ref{WorkAndResults}.

Even though personas have become a popular method for raising awareness of users'
needs in a project, it is important to remember that personas are \emph{not} a
replacement for actual users. That is, one should not create personas out of thin air to replace gathering
input from users. This may allow stakeholders to think about the
users, but it will be problematic when creating personas with disabilities. This is because
many people have misconceptions about how people with disabilities interact with
technology and this can lead to these personas having extra powers or disadvantages that
they may not have. As pointed out by Grudin and Pruitt \cite{Grudin2002} personas can be used poorly and for most people ``\ldots a more solid foundation will prove necessary.'' In particular, Grudin and Pruitt recommend basing personas on user research and facts. 

\section[Methodology]{Methodology}
As mentioned in the Section \ref{SOTA}, even though personas are fictional, they should be based on experiences with and
information from real users \cite{Calabria2004}. We underscore that personas cannot replace contact with
real users altogether, but rather be used as a supplement, and as a way of keeping a
continual focus on the users throughout the project life-cycle. There are many ways that
one can go about collecting information about real users and, depending on the resources
available, selecting more than one method may be useful.

One way that can be useful for collecting information about users is by simply asking
them. This includes methods like using focus groups, interviews, and surveys. Observation
is another good method. As Gould \cite{Gould1995} points out, you may not have any idea
about what you need to know about users and their environment until you see them. It is
useful to study information from case studies and other user research. Market information may be another
source to consider including.

When it comes to recruiting people with disabilities, having contacts inside the user
organizations that support people with disabilities is a good start. User organizations
can contact members for you and provide opportunities for you to talk to users at meetings
or provide a location to host a focus group: a well-known location can make it easier for
people with visual or physical impairments to participate rather than traveling to your site, which is unlikely unknown to them.
Using surveys can also be a way of gathering information. An online survey can help you
reach a wider audience that might have been impossible or cost prohibitive otherwise, but
it is important that the tools for gathering the information are usable for your audience.
For example, using a web survey tool may create a survey that is inaccessible to people
using certain types of AT like screen readers \cite{Wentz2009}. A plain text email with
the questions may be an alternative for getting these voices heard. Getting more
information will help create more well-rounded personas and highlight different issues
that will need to be taken into a system.

Looking at the AT used by people with disabilities also helps in creating personas with
disabilities. Some personas will be using AT for accessing information. It is important to
know how these technologies work and how people work with them. It is vital that someone in
the design team has actual experience working with people with disabilities, either from
user tests or from teaching them to use technology. You should at least include people
with this kind of experience in the process of creating personas. One way to do this could
be to invite them to a persona workshop.

As an example of how to involve users, in one project, the UNIMOD project \cite{UNIMOD2007}, a navigation system for drivers
working at a rehabilitation center was to be developed. A persona workshop was arranged on
the premises of the rehabilitation company. Employees at the company with ample
experience with the target population were invaluable during the persona workshop. As
various aspects of the target population were discussed during the persona creation
process, the employees could fill in with related real-life stories. The stories were
told in connection to discussions of traits, needs, attitudes, and habits of the various
personas. Later in the project, project participants remembered several of these stories;
they were referred to when using the personas, and they were useful for keeping the
personas alive.

As outlined by Pruitt and Adlin \cite{Pruitt2006}, a persona workshop gathers the stakeholders to
generate personas based on assumptions and factoids. Assumptions are quotes, opinions, or
goals that these potential personas would have along with a possible name for the persona.
Factoids are small facts or news items from literature, research, or data from your own
user research. During the workshop, participants re-read through the collected research and writes out factoids (e.g., on post-it notes). This can also be repeated for assumptions.
Starting first with the assumptions, stakeholders build groups of similar assumptions to
see if there are any patterns that emerge. This could be done digitally with mind-mapping software or in analog with post-it notes and a clear wall or whiteboard.  This process is repeated with the factoids
usually resulting in a rearrangement or new groups being created. These groups are the
starting point for creating persona skeletons.

Persona skeletons are the outlines of the actual personas. They consist of the assumptions
and factoids that were collected earlier, but they also are where sketches of information
about the personas start to emerge. One way of organizing this information is to use a
template with all the different areas of the final persona description. Start by filling
in information first as keywords and continue until you have fleshed-out the entire
section. A mind map is another good way of creating the ``bones'' of the skeletons before
adding ``flesh.''

Once everyone agrees on the persona skeletons, writing up of the actual personas can
begin. The outcome is usually what most people see when they are presented personas, the
persona description as detailed in Section \ref{SOTA}. If the persona has a disability, this information is
also presented along with information about the AT the persona uses. Since others in the
project may not have an understanding about how a person with a disability works with an
AT, it may be necessary to include information about how a disability affects a persona or
how particular AT plays a part in the persona's life. After this, the personas are ready to
become active participants in the project.

How many personas should be created for a project? If we want to
aim for universal design, targeting the four main groups of
disabilities is a good start. That is, create personas with vision, hearing, movement, and cognitive
impairments. Yet, as mentioned in Section \ref{SOTA}, one should keep in mind that each of these impairments group are 
diverse and have different abilities. Another option to consider is to create an
elderly persona. Elderly personas usually have a combination of several milder versions of
impairments from these groups. In our experience, we have found that three to six personas
is a manageable amount of work and covers important aspects of our target groups.

\section[Work and Results]{Work and Results}
\label{WorkAndResults}
This technique has been used in several of our projects. Currently we are using it in
researching the Internet of Things (IoT) and Ambient Assisted Living (AAL). We wanted to
ensure that the needs of users with disabilities were included in the requirements and
prototypes. For the IoT project, we wanted to examine the issues that people with vision
impairment and those with dyslexia have when interacting with the Internet of Things. Of
our five personas, one has twenty percent vision, another has dyslexia, and another is
elderly and has begun to suffer from mild dementia. We have documented the different AT
these personas use and tried to describe real issues. For example, our persona with vision
impairment uses a screen reader and magnifier, but has one version at work and another at
home; the different software results in our persona sometimes forgetting which keyboards
shortcuts work where.

The AAL project focuses on elderly people's use of mobile phones and getting help on them.
We want to make sure that we can reach the largest group of the elderly as possible. All
the personas for this project have a slight vision impairment and other disabilities like
hearing loss or problems remembering information. Since the project is about using mobile
phones and asking for assistance, we made sure that the elderly personas have similar
attitudes to technology and to learning new information that match the different focus
groups we held when gathering user requirements. This is also reflected in our personas
choice of mobile phone.

We have found that including disabilities in the persona creation phase has helped in
raising awareness for universal design and accessibility both during the creation process
and in many other areas of the project. One of the most obvious places was in the creation
of the user scenarios. Our personas became the performers in these scenarios and it was
necessary to ensure that the different actions in the scenarios could be accomplished by
the specific persona. This bubbled up into later requirements work such as selecting
technology, defining use cases, and in recruiting informants for evaluations. It is
important to keep the personas in project participants' thoughts. This has been done in
different ways. Each month we get an email message with a story from one of the personas explaining an issue
that persona faced with technology or some other aspect that is related to the project.
The task of writing such a story is distributed among the project participants. During the
process of creating these stories and describing in detail how a persona interacts with the
technology may raise questions for the story's author. Is the story realistic for the actual persona? Would the
persona actually do it in this way? If the project participant authoring the story does not
have experience with how people with the particular type of disability interacts with
technology, the story should be presented to someone who has this experience, or even
users themselves. The process of writing the story and getting it validated either by
experts or by users, helps to reveal potentially wrong assumptions among the project
participants. Because the process is creative and active, it encourages learning about the
issues this persona, and people with similar disabilities, have. Project participants also received gifts related to the personas, such as a chocolate
Advent Calendar with their pictures, reminding project participants about the personas
every day in December. Pruitt and Grudin \cite{Pruitt2003} list many additional ideas that can keep
personas alive.

Another valuable method to utilize the personas is to do \emph{persona testing} with
prototypes as an analog to user testing. Tasks for the personas to perform are created as
in a user test. The personas are divided between members of the project team according to
their experience and familiarity with the disability that the persona has. Then, the team
member acts as the persona while doing the tasks with the prototype. The more experience
the team member has, either from user tests with or training people with the type of
disability that the persona has, the easier it is to do a realistic and credible acting
performance when persona testing the prototype. If none of the team members have this
experience, one should consider inviting someone who does. The person performing the
persona test may take notes, but we advise to have another team member to be observer and
to takes notes during the persona testing. This approach is informal and relatively quick
to do. It can be done in between user testing with real users. We have also used persona
testing to pilot user tests, to identify potential problems that can be corrected before
the user test, and to see how many and what types of tasks would be fruitful to do in the
user test.

\section{Impact} 
As more countries have started to create requirements that new ICT
targeting the public be universally designed or accessible, including the needs of people
with disabilities will be increasingly needed for projects. The methodology outlined above
is useful for others that want to include the perspectives of people with disabilities in
their work. It requires some initial work upfront to build competence and knowledge about
AT and people with disabilities, but this work would be needed in any sort of work for
universal design. Once this knowledge is acquired, it can easily be incorporated into any
persona creation process. Rather than using personas to replace user research, it can be used as a means to elicit knowledge and experience from people in your team or network that do have experience with people who have the type of disability the persona has.

\section{Conclusion}
We have presented the state of the art for persona creation and outlined a methodology for
creating personas with disabilities. In our own work, we have found that using this
methodology has helped raise awareness among partners about the needs of people with
disabilities and has ensured that the personas' needs are included in all steps in the
project. We hope that this methodology will result in more universally designed ICT and
that others will use this technique themselves. We also have found that it is important
that to involve people in the project who have experience with how people with
disabilities use technology. This can either be with people with disabilities themselves
or others who aid people with disabilities or research
issues in the universal design of ICT. Including these people can only help ensure that a
project focuses on the needs for universal design.

Finally, it is not sufficient to simply create personas. They need to be used in order for
them to be alive. This can include things like creating stories to document things that
are happening in a persona's life and remind everyone that they should keep these personas
in mind in the work that they do. A persona walkthrough using the proper AT is also a
concrete way to remind everyone about what type of users will actually be using the final
product or service. Following this advice should ensure that personas you create will
capture the needs of people with disabilities and capture the attention
of the project members.

\subsubsection*{Acknowledgments.}
This research is funded as part of the uTRUSTit project. The uTRUSTit project is funded by 
the EU FP7 program (Grant agreement no: 258360).

\bibliography{ICCHP-Personas}
\bibliographystyle{splncs}

\bigskip
\end{document}